\documentclass[ALICE,manyauthors]{cernphprep}

\usepackage{hyperref}
\usepackage[sort, numbers]{natbib}
\usepackage{multirow}
\usepackage{breakurl}

\newcommand {\pT}{\ensuremath{p_{\mathrm{T}}} }
\newcommand {\PbPb}{\mbox{Pb--Pb} }
\newcommand {\gmom} {\mbox{\rm GeV$\kern-0.15em /\kern-0.12em c$}}
\newcommand {\tev} {\mbox{${\rm TeV}$}}
\newcommand{\Kzs}{\mbox{$\mathrm {K^0_S}$}}
\newcommand{\rmLambda}{\mbox{$\mathrm {\Lambda}$}}
\newcommand{\sqrtsNN} {\mbox{$\sqrt{s_\mathrm{NN}}$}}
\newcommand {\pseudorap} {\mbox{$\eta$}}
\newcommand{\Nev}{13}
\newcommand {\dNdy}{\mathrm{d}N/\mathrm{d}y }
\newcommand {\cm} {\mbox{${\rm cm}$}}

\begin{document}%
%
\begin{titlepage}
\PHnumber{2013-132}                 
\PHdate{July 18, 2013}              
%
%
\title{\Kzs\ and \rmLambda\ production in \PbPb collisions at 
\sqrtsNN~=~2.76~$\tev$}
\ShortTitle{\Kzs\ and \rmLambda\ production}   
%
\Collaboration{ALICE Collaboration%
         \thanks{See Appendix~\ref{app:collab} for the list of collaboration
                      members}}
\ShortAuthor{ALICE Collaboration}      
\begin{abstract}
The ALICE measurement of $\Kzs$ and $\rmLambda$ production  
at mid-rapidity in \PbPb collisions at $\sqrtsNN = 2.76~\tev$ is presented.
The transverse momentum ($\pT$) spectra are shown for several
collision centrality intervals
and in the $\pT$
 range from $0.4~\gmom$ ($0.6~\gmom$ for \rmLambda) to $12~\gmom$. 
The \pT dependence of the  $\rmLambda/\Kzs$ ratios 
exhibits maxima in the vicinity of 3~\gmom, and the positions 
of the maxima shift towards higher \pT with
increasing collision centrality. 
The magnitude of these maxima
increases by almost a factor of three between most peripheral
and most central \PbPb collisions.  
This baryon excess at intermediate \pT is not observed
in pp interactions at $\sqrt{s} =0.9~\tev$ 
and at $\sqrt{s} =7~\tev$. 
Qualitatively, the baryon enhancement in heavy-ion collisions is expected 
from radial flow. However, the measured $\pT$ spectra  above 2~\gmom\ 
progressively decouple from hydrodynamical-model calculations. 
For higher values of $\pT$,  models that incorporate the influence of 
the medium on the fragmentation and hadronization processes
describe qualitatively the $\pT$ dependence of the $\rmLambda/\Kzs$ ratio.

\end{abstract}
\end{titlepage}

\setcounter{page}{2}
%
Collisions of heavy nuclei at ultra-relativistic energies are used to investigate
a deconfined high temperature and density state of nuclear matter, 
the quark--gluon plasma. 
The transverse momentum ($\pT$) spectra of identified hadrons and their ratios 
provide a means for 
studying the properties of this state of matter and the 
mechanisms that transform quasi-free partons into observed hadrons. 
It was observed at the Relativistic Heavy Ion Collider 
(RHIC)~\cite{Adler:2003rr, Adams:2006wk}, 
that the $\rmLambda/\Kzs$ and p$/\pi$ ratios at intermediate $\pT$ 
(from about 2 to about 6~\gmom) are markedly 
enhanced in central heavy-ion collisions when compared with the peripheral
or pp results. A similar observation was also made at the Super Proton
Synchrotron~\cite{Schuster:2006jt}. 
These observations led to a revival and further development of
models based on the premise that deconfinement opens an additional mechanism
for hadronization by allowing two or three soft quarks from the bulk to combine
forming a meson or a baryon~\cite{Fries:2003fr, Fries:2008bh}.
The baryons then appear at a higher
\pT than the mesons, since their momentum is the sum of the momenta of 
three quarks, instead of only two.
If the (anti-)quarks generated by (mini)jet fragmentation are also involved in 
recombination~\cite{Hwa:2004ng}, the baryon enhancement could extend 
to even higher momenta, up to 10--20~\gmom~\cite{Hwa:2008qi}.
At lower $\pT$, the hydrodynamical radial flow also contributes to the 
baryon enhancement, because the baryons, being heavier, are pushed to higher $\pT$ 
than the mesons. But the applicability of such models is limited 
to transverse momenta below 2~\gmom, above which the observed $\pT$ 
spectra start to deviate from hydrodynamical calculations.

The evolution of the baryon to meson ratio with
collision energy, comparisons with pp events and a study of the centrality
dependence in nucleus--nucleus collisions
 provides additional information about this 
``baryon anomaly''~\cite{Lamont:2007ce}.
In Pb--Pb collisions at the Large Hadron Collider (LHC) energies, 
that are around 14 times higher than 
those at RHIC, the maximum of the $\rmLambda/\Kzs$ ratio is expected to be shifted towards higher
$\pT$, because of an increased partonic radial 
flow~\cite{Fries:2003fr, Fries:2008bh}.
In contrast, the $\rmLambda/\Kzs$ ratio measured
in elementary pp collisions should not change significantly with the 
center-of-mass energy, since the particle production is presumably 
dominated by fragmentation processes. 

The relative contribution of different hadronization 
mechanisms changes with hadron momentum. 
While at intermediate \pT
recombination might be dominating, fragmentation could take
over at higher $\pT$, depending on the underlying momentum
distributions of the quarks.
For this reason it is important to identify baryons and mesons in a wide momentum range.
The topological
decay reconstruction of $\Kzs$ and $\rmLambda$ provides an opportunity
to extend the baryon and meson identification from low to high transverse
momenta, which can not easily be achieved using other particle
identification methods without introducing additional systematic effects. 

In this Letter we
present the $\Kzs$ and $\rmLambda$ transverse-momentum spectra and the 
$\rmLambda/\Kzs$ ratios from \PbPb
collisions at $\sqrtsNN=2.76$~TeV 
recorded in November 2010's heavy-ion run of the LHC.  
The $\pT$ dependence of the $\rmLambda/\Kzs$ ratios is compared with
pp results obtained at $\sqrt{s} =  0.9$ and 7~TeV, that bracket
the Pb--Pb measurements in energy.

A description of the ALICE apparatus can be found in~\cite{Aamodt:2008ul}.
For the analysis presented here, we used the Time Projection Chamber (TPC)
and the
Inner Tracking System (ITS) to reconstruct
charged particle tracks within the  pseudo-rapidity interval of
$|\pseudorap| <0.9$. Particle momenta were determined from the track curvature in a
magnetic field of 0.5~T. 
The two VZERO scintillator counters, covering pseudo-rapidity ranges of  $2.8
< \pseudorap < 5.1$ (VZERO-A) 
and $-3.7 < \pseudorap < -1.7$ (VZERO-C), provided a signal proportional to
the number of charged particles in these acceptance regions.
The VZERO detectors together with the two innermost Silicon Pixel Detector 
(SPD) layers of the ITS, 
positioned at radii of 3.9 and 7.6 cm (acceptance $|\eta|<2.0$ and
$|\eta|<1.4$ respectively), were used as an interaction trigger.
To select a pure sample of hadronic interactions, only events with at least one particle hit in each of the three
trigger detectors (SPD, VZERO-A and VZERO-C) 
were accepted offline.
The selected events were required to have reconstructed primary 
vertices with a position along the beam direction within $\pm 10~\cm$
of the nominal center of the detector to ensure a uniform acceptance in
pseudo-rapidity for
the particles under study.  The events were then classified
according to the collision centrality, based on the sum of the amplitudes in the VZERO
counters fitted with a Glauber model description of the collisions,
as discussed in~\cite{Abelev:2013qoq}. After these selections,
we retained for the final analysis \Nev~million events in the collision
centrality range from 0 to 90\% of the nuclear cross-section.  

%
The weakly decaying neutral hadrons (\Kzs\ and \rmLambda) were reconstructed using
their distinctive V-shaped decay topology in the channels (and branching ratios) $\Kzs\rightarrow\pi^+\pi^-$ (69.2\%) and \rmLambda$\rightarrow\mathrm{p}\pi^-$ (63.9\%)~\cite{RPP}.
The reconstruction method forms so-called V0 decay candidates and the details are described in~\cite{Aamodt:2011zza}. 
Because of the large combinatorial background in
Pb--Pb collisions, a number of topological selections had to be more 
restrictive than those used in the pp analysis.
In particular,  the cuts on the minimum distance of closest approach between the
V0 decay products and 
on the minimum cosine of the V0 pointing angle (the angle between the line
connecting the primary and V0 vertices and the V0
momentum vector)~\cite{Aamodt:2011zza} were changed to 
one standard deviation  and to 0.998, respectively. 
In addition, we retained  only the V0 candidates reconstructed
in a rapidity window of $|y|<0.5$, with their decay-product
tracks within the acceptance window $|\eta|<0.8$. To further suppress the
background, 
we kept only V0 candidates satisfying the cut on the proper decay length 
$l_\mathrm{T}\cdot m/\pT < 3~c\tau~(4~c\tau)$,
where $l_\mathrm{T}$ and $m$ are the V0 transverse decay length and 
nominal $\rmLambda$ ($\Kzs$) mass~\cite{RPP}, and $c\tau$
is  7.89~cm (2.68~cm) for  $\rmLambda$ ($\Kzs$)~\cite{RPP}. 
For the \rmLambda\ candidates with $\pT<1.2$~\gmom, a conservative 
three-standard-deviation particle-identification cut
on the difference between the specific energy loss (d$E$/d$x$) 
measured in the TPC and the expected energy loss as defined
by a momentum-dependent parameterization of the Bethe--Bloch
curve was applied for the proton decay-product tracks. 
To reduce the contamination of \rmLambda\ reconstructed as \Kzs, an
additional selection was applied in the Armenteros--Podolanski 
variables~\cite{armenteros} of  \Kzs\
candidates, rejecting candidates with $\pT^\mathrm{arm} <
0.2\times|\alpha^\mathrm{arm}|$. 
Here, $\pT^\mathrm{arm}$ 
is the projection of the positively 
 (or negatively) charged decay-product momentum on the plane perpendicular to
the V0 momentum.
The decay asymmetry parameter $\alpha^\mathrm{arm}$ is defined as
$\alpha^\mathrm{arm}=(p^{+}_\parallel-p^{-}_\parallel)/(p^{+}_\parallel+p^{-}_\parallel)$,
 where $p^{+}_\parallel(p^{-}_\parallel)$
 is the projection of the positively
 (negatively) charged decay-product momentum on the momentum of the V0.
The minimal radius of the fiducial volume of the secondary
vertex reconstruction was chosen to be 5~cm to minimize systematic
effects introduced by efficiency corrections. It was verified that the 
decay-length distributions reconstructed within this volume were
exponential and agreed with the $c\tau$ values given in the
literature~\cite{RPP}.

\begin{figure}
\centering
\resizebox{0.5\textwidth}{!}{%
\includegraphics{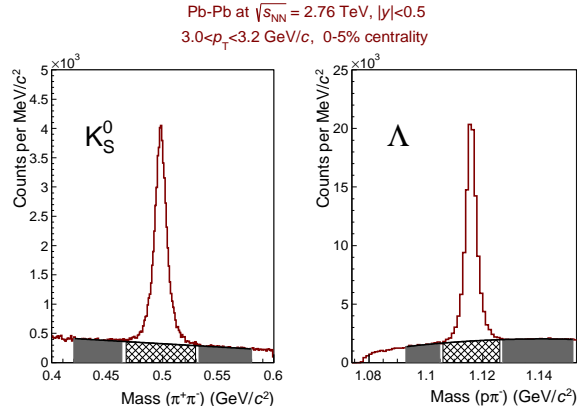}
}
\caption{Examples of invariant mass distributions for 
$\Kzs$ and $\rmLambda$. 
The filled areas
to the sides of the peaks were used to fit the background in order to
estimate the background level under the peaks, indicated as the light
shaded areas. 
\label{fig:mass}
}
\end{figure}

The raw yield in each $\pT$ bin was extracted from the invariant-mass 
distribution obtained for this momentum bin.
Examples of such distributions are shown in Fig.~\ref{fig:mass}.
The raw yield was calculated by subtracting a fit to the background from 
the total number of V0 candidates in the peak region. 
This region was $\pm 5\sigma$ for $\Kzs$, and $\pm(3.5\sigma + 2$~MeV/$c^2)$
(to better account for tails in the mass distribution at low $\pT$) 
for $\rmLambda$.
The $\sigma$ was obtained by a Gaussian fit to the mass peaks.
The background was determined by fitting polynomials of first or second
order to side-band regions left and right of the peak region.

The overall reconstruction efficiency corrections were extracted from a 
procedure based
on \textsc{HIJING} events~\cite{Gyulassy:1994ew}
and using \textsc{GEANT3}~\cite{:1994zzo} for transporting simulated
particles, followed by a full calculation of the detector responses and
reconstruction done with the ALICE simulation and reconstruction framework~\cite{aliroot-tdr}. 
The estimated efficiency included the geometrical acceptance of the
detectors, 
track reconstruction efficiency, the efficiency of the applied topological
selection cuts, and the  branching ratios for the V0 decays. 
The typical efficiencies for both particles were about 30\% for $\pT>4$~\gmom,
dropping to 0 at $\pT\sim 0.3$~\gmom. 
The efficiencies did not change with the event centrality for $\pT$
above a few~\gmom.
However, at lower $\pT$, they were found to be dependent on the event 
centrality. For $\Lambda$ at  $\pT<0.9$~GeV/$c$ the difference is about
factor 2 
between the 0--5\% and 80--90\% centrality intervals.
This was because the distributions of the topological variables used in the
selections were changing with the centrality, whereas the corresponding 
threshold cut values were kept constant. The effect was well reproduced 
by the Monte Carlo simulations.
The final momentum spectra were therefore corrected in each centrality bin 
separately.

The spectra of \rmLambda\ were in addition corrected for 
the feed-down contribution
coming from the weak decays of $\Xi^-$ and $\Xi^0$. 
For this purpose, a two-dimensional response matrix, correlating the \pT 
of the detected
decay \rmLambda\ with the \pT of the decayed $\Xi$, was generated from 
Monte-Carlo simulations. By normalizing this matrix to the measured
$\Xi^-$ spectra~\cite{xi_omega}, the distributions of the  feed-down 
\rmLambda\ were
determined and subtracted from the inclusive \rmLambda\ spectra. 
The phase space distribution and total yield for 
the $\Xi^0$ were assumed to be the same as  
for the $\Xi^-$.
The feed-down correction thus obtained was found to be a smooth
function of $\pT$ with a maximum of about 23\% at $\pT\sim 1$~\gmom\ 
and monotonically decreasing to 0\% at $\pT>12$~\gmom. 
As a function of centrality, this correction changed by only a few per 
cent. 

Since the ratio $\Xi^-/\Omega^-$
in Pb--Pb collisions at  $\sqrtsNN=2.76$~TeV was measured to be 
about 6~\cite{nicassio_sqm}, and taking into account that the
branching ratio $\Omega^-\rightarrow\rmLambda$K$^-$ is 67.8\%~\cite{RPP}, 
the feed-down
contribution from decays of $\Omega^-$ baryons would be about 1\%, which 
is negligible compared with other sources of uncertainty (see below).
Also, we did not correct the $\rmLambda$ spectra for the feed-down from
non-weak decays of $\Sigma^{0}$ and $\Sigma(1385)$ family.

The fraction of \rmLambda's produced
in hadronic interactions with the detector material was estimated using 
the detailed Monte Carlo simulations mentioned above. Since this fraction
was found to be less than 1\%, it was neglected.

\begin{figure*}
\resizebox{0.98\textwidth}{!}{%
\includegraphics{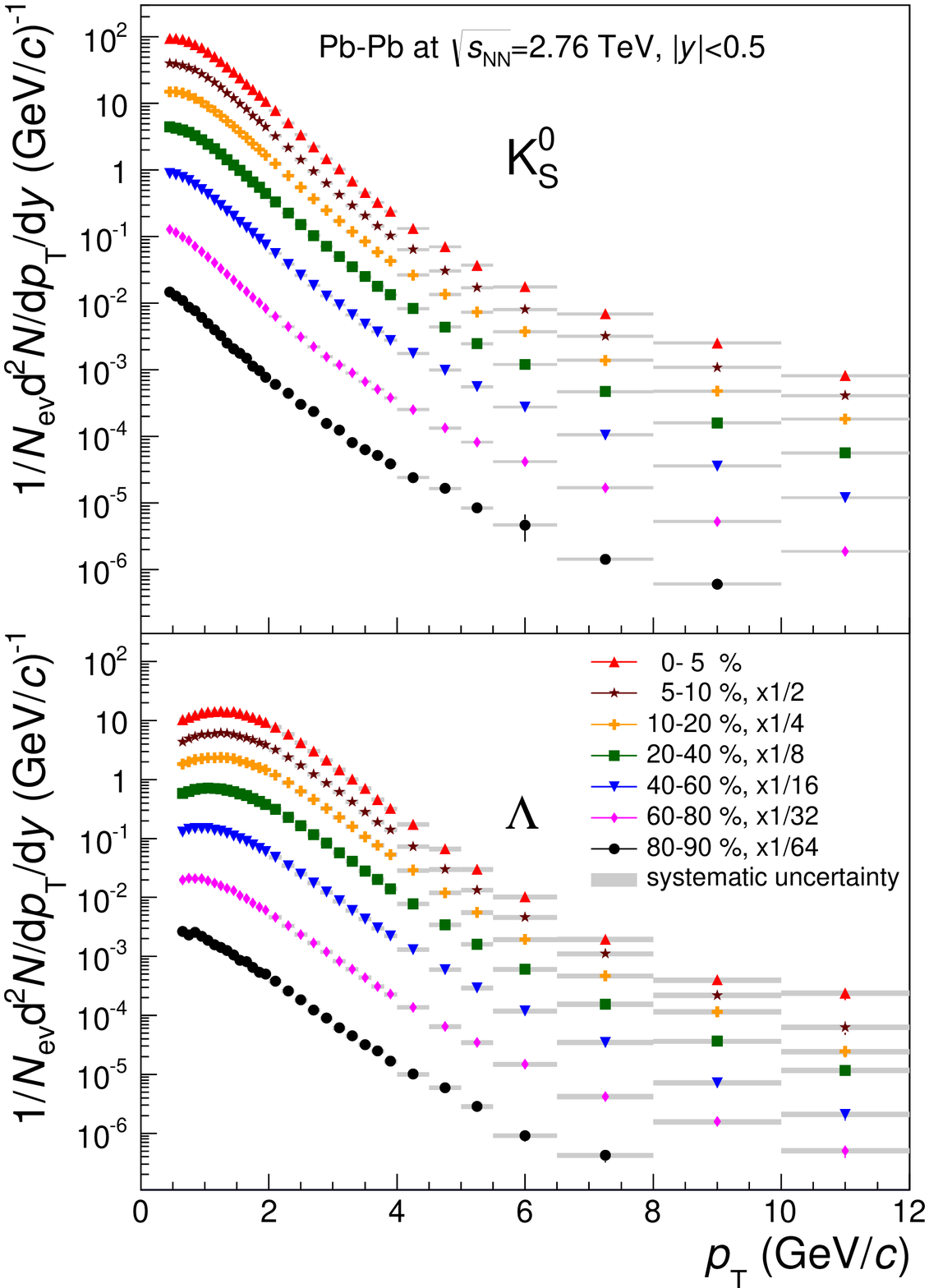}
\includegraphics{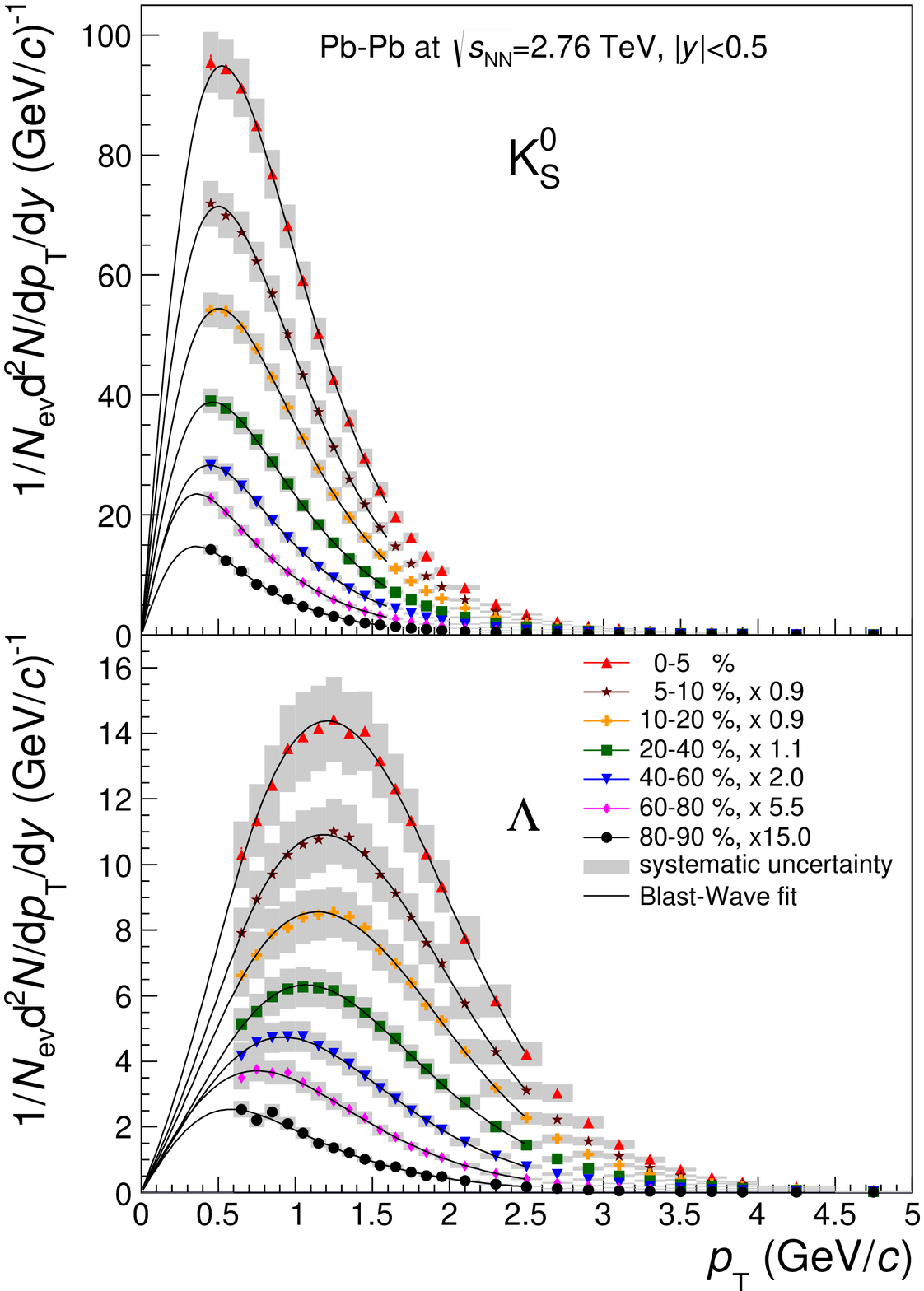}
}
\caption{$\Kzs$ and $\rmLambda$ transverse momentum
  spectra for different event centrality intervals shown in logarithmic
  (left) and linear (right) scale. 
The curves represent results of blast-wave fits~\cite{Schnedermann:1993ws}.
\label{fig:spectra}
}
\end{figure*}

The following main sources of systematic uncertainty were considered:
raw yield extraction, feed-down, efficiency corrections, and the uncertainty
on the amount of crossed material.
These were added in quadrature to yield the overall systematic uncertainty on the \pT
spectra for all centralities.

The systematic uncertainties on the raw yields were estimated by using 
different functional shapes
for the background and by varying the fitting range. 
Over the considered momentum range, 
the obtained raw yields varied within 3\%
for \Kzs\ and 4--7\% for \rmLambda. 

As a measure for the systematic
uncertainty of the feed-down correction, we used the spread
of the values determined for different centrality ranges with respect 
to the feed-down correction estimated for minimum bias events. 
This deviation was found
to be about 5\% relative to the overall \rmLambda~yield. 
 
The systematic uncertainty associated with the efficiency correction was 
evaluated by varying one-by-one the topological, track-selection and PID cuts.
The cut variations were chosen such that the extracted uncorrected yield of the $\Kzs$ and $\rmLambda$ 
would change by 10\%. To measure the systematic uncertainty related to each
cut, we used as a reference the corrected spectrum obtained with the
nominal cut values. 
For \rmLambda, 
the feed-down correction was re-evaluated and taken into account for every 
variation of the cut on the cosine of the pointing angle. 
The overall $\pT$-dependent
systematic uncertainty associated with the efficiency correction was then
estimated by choosing the maximal (over all cut variations) 
deviation between varied and nominal spectra values
obtained in each momentum bin. For the momentum range considered,
this systematic uncertainty was determined to be 4--6\% for both \Kzs\ and
 \rmLambda.

The systematic uncertainty introduced because of possible imperfection
in the description of detector material in the simulations was
estimated in~\cite{Aamodt:2011zza} and amounted to 1.1--1.5\% for $\Kzs$ 
and 1.6--3.4\% for $\rmLambda$.

Since the systematic uncertainties related to the efficiency correction
are correlated for the $\rmLambda$~and \Kzs~spectra, they
partially cancel in the $\rmLambda/\Kzs$ ratios. 
These uncertainties were evaluated by dividing $\rmLambda$~and \Kzs~spectra obtained with the same cut variations and found to be
half the size of those that would be obtained if the uncertainties of the
$\rmLambda$~and \Kzs~spectra were assumed to be uncorrelated.  
Altogether, over the considered momentum range, the maximal systematic 
uncertainty for the measured  $\rmLambda/\Kzs$ 
ratios was found to be about 10\%.

\begin{figure*}
\resizebox{0.98\textwidth}{!}{
\includegraphics{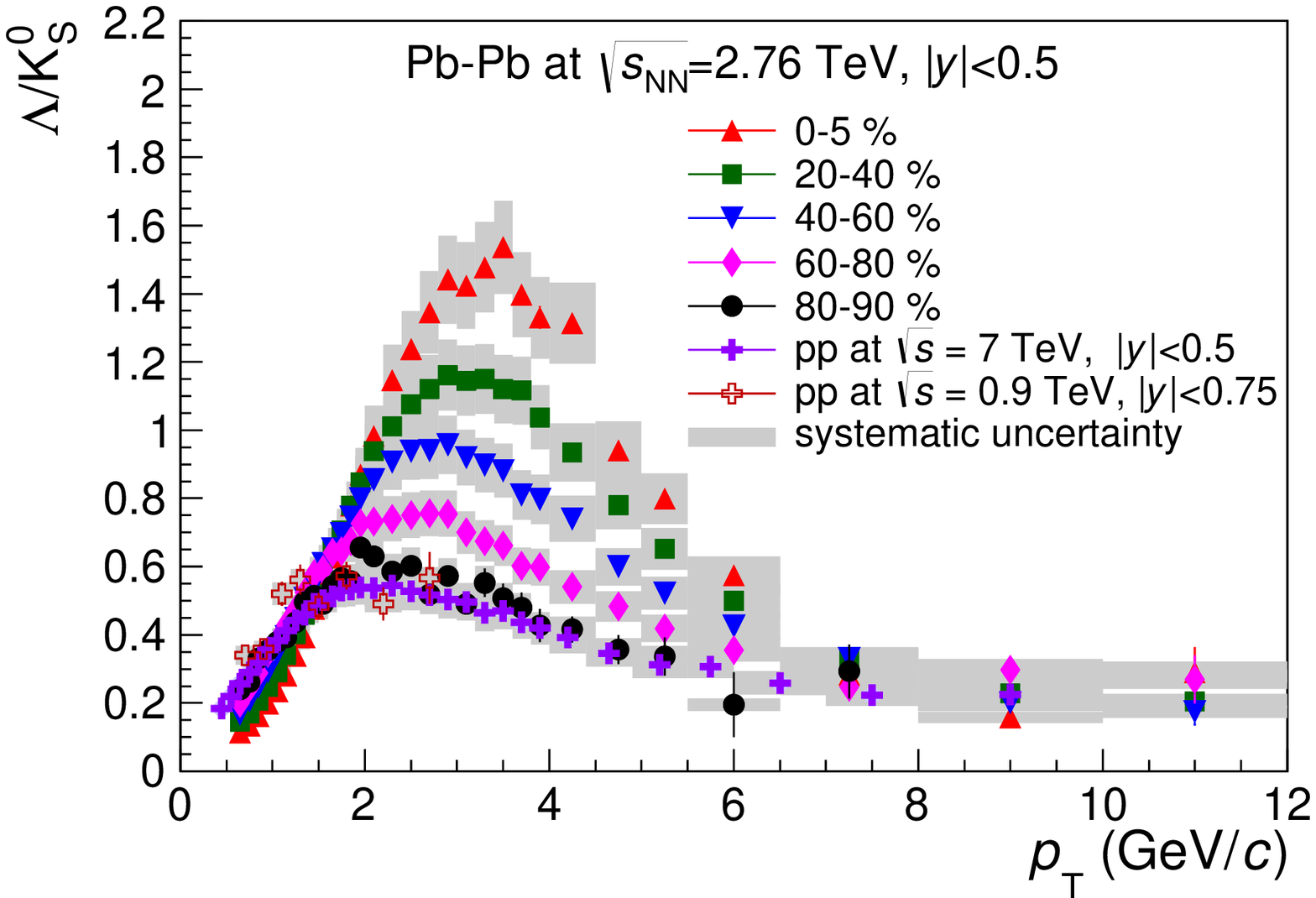}
\includegraphics{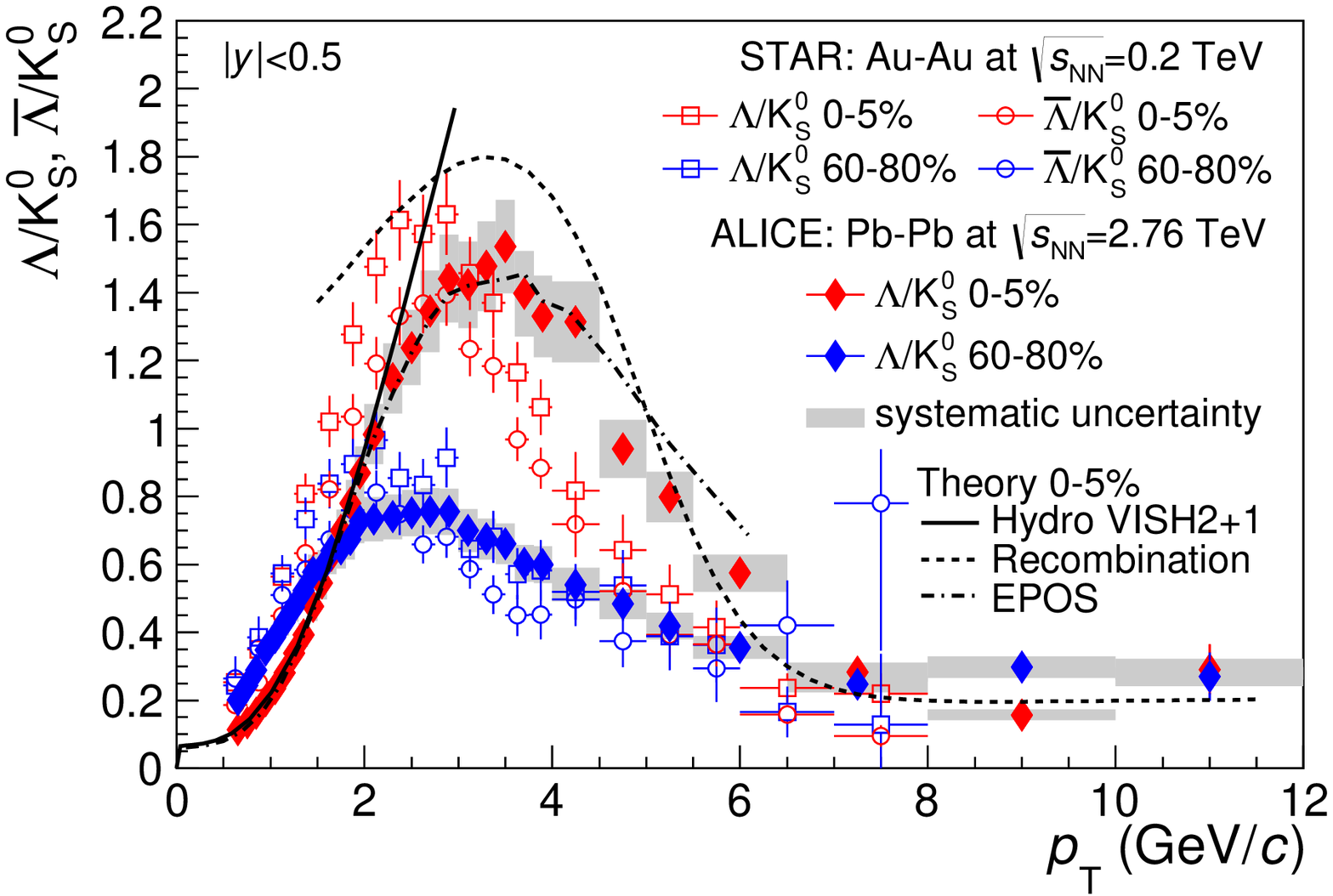}
}
\caption{Left: $\rmLambda/\Kzs$ ratios as a function of $\pT$ for different 
event centrality intervals in Pb--Pb collisions at 
$\sqrt{s_\mathrm{NN}} = 2.76$~TeV and pp collisions at $\sqrt{s} = 0.9$~\cite{Aamodt:2011zza} 
and 7~TeV~\cite{pp7TeV}. 
Right: Selected $\rmLambda/\Kzs$ ratios as a function of $\pT$ compared
with $\rmLambda/\Kzs$ and $\bar{\rmLambda}/\Kzs$ ratios measured in Au--Au collisions at
$\sqrt{s_\mathrm{NN}} = 200$~GeV~\cite{Agakishiev:2011ar}.
The solid, dashed and dot-dashed lines show the
corresponding ratios from a hydrodynamical 
model~\cite{Song:2007fn,Song:2007ux,Song:2008si}, 
a recombination model~\cite{Song:2011qa} and the EPOS model~\cite{EPOS}, respectively.
\label{fig:ratio}
}
\end{figure*}

The transverse-momentum spectra of $\Kzs$ obtained in different centrality 
intervals 
were compared with the spectra of charged kaons also measured by
ALICE~\cite{piKpPbPb:2013}. The two sets of spectra agree within the 
systematic uncertainties.

%
%
The corrected \pT spectra are shown
in logarithmic scale in 
Fig.~\ref{fig:spectra} (left). The spectra were fitted using the blast-wave
parameterization described in~\cite{Schnedermann:1993ws}. 
The resulting curves are superimposed in Fig.~\ref{fig:spectra}
(right), with a linear scale and for a restricted momentum range, 
to emphasize the low-$\pT$ region. The fit range in \pT was from the lowest 
measured point up to 2.5~\gmom\ (1.6~\gmom)
for \rmLambda\ (\Kzs).
The fitting functions were used to extrapolate the spectra to zero
\pT to extract integrated particle yields $\dNdy$. 
The results are given
in Table~\ref{tab:fits}. 
The systematic uncertainties of the integrated yields 
were determined by shifting the data points of the spectra 
simultaneously within
 their individual systematic uncertainties and reapplying the
fitting 
and integration procedure. In addition, an extrapolation uncertainty was
estimated, by using alternative (polynomial, exponential and 
L\'{e}vy-Tsallis~\cite{Tsallis:1988fk, Abelev:2006cs}) functions 
fitted to the low-momentum part of the spectrum, 
and the corresponding difference in obtained values was added in quadrature.

The $\pT$ dependence of the $\rmLambda/\Kzs$ ratios, formed for each
centrality interval by a division
of the respective measured $\pT$ spectra, 
is presented in Fig.~\ref{fig:ratio} (left panel).
For comparison, the same ratios measured in minimum bias pp collisions at
$\sqrt{s}=0.9$~\cite{Aamodt:2011zza} and $7~\tev$~\cite{pp7TeV}  are
plotted as well.

\begin{table*}[tbhp]
\centering
\caption{\label{tab:fits}
Integrated yields, $\dNdy$, for $\rmLambda$ and $\Kzs$ with uncertainties 
which are dominantly systematic. A blast-wave fit is used to extrapolate 
to zero $\pT$. 
Fractions of extrapolated yield are specified.
Ratios of integrated yields, $\rmLambda /\Kzs$, for each centrality bin
with the total uncertainty, mainly from systematic sources, are shown.}
\resizebox{\textwidth}{!}{%
\begin{tabular}{@{} cc|ccccccc @{}} 
\hline
\hline						
&  & 0--5\% & 5--10\% & 10--20\% & 20--40\% & 40--60\% & 60--80\% & 80--90\%  \\
\hline
\multirow{2}{*}{\rmLambda~~} & $\dNdy$ & $26\pm3$ & $22\pm2$ & $17\pm2$ & $10\pm1$ & $3.8\pm0.4$ & $1.0\pm0.1$ & $0.21\pm0.03$ \\
& $\pT < 0.6$~\gmom~frac.~ & 10\% & 11\% & 12\% & 14\% & 18\% & 24\% & 32\% \\
\hline
\multirow{2}{*}{$\Kzs~~$} & $\dNdy$ & $110\pm10$ & $90\pm6$ & $68\pm5$ & $39\pm3$ & $14\pm1$ & $3.9\pm0.2$ & $0.85\pm0.09$ \\
& $\pT < 0.4$~\gmom~frac.~ & 20\% & 21\% & 21\% & 23\% & 25\% & 31\% & 33\% \\
\hline
 & Ratio $\dNdy$ $\rmLambda /\Kzs$~ & ~$0.24 \pm 0.02$~ &  ~$0.24 \pm 0.02$~ & ~$0.25 \pm 0.02$~ & ~$0.25 \pm 0.02$~  & ~$0.26 \pm 0.03$~   & ~$0.25 \pm 0.02$~    & ~$0.25 \pm 0.02$~   \\

\hline
\hline						
\end{tabular}
}
\end{table*}


The $\rmLambda /\Kzs$ ratios observed in pp events 
at $\sqrt{s} = 0.9$ and 7~TeV agree within uncertainties over the presented
$\pT$ range, and they bound in energy the \PbPb 
results reported here. The ratio measured in the most peripheral 
\PbPb collisions is compatible with the pp measurement, where there is a
maximum of about 0.55 at $\pT\sim 2$~\gmom.  
As the centrality of the \PbPb collisions increases, the maximum value of
the ratio also increases and its position shifts towards higher momenta.
The ratio peaks at a value of about 1.6 at $\pT\sim 3.2$~\gmom\ for the 
most central Pb--Pb collisions. 
This observation may be contrasted to the ratio of the integrated 
$\rmLambda$ and $\Kzs$ yields which does not change with centrality 
(Table~\ref{tab:fits}).
At momenta above $\pT\sim7$~\gmom, the $\Lambda$/\Kzs\ ratio is
independent 
of collision centrality and $\pT$, within the uncertainties, 
and compatible with that measured
in pp events.

A comparison with similar measurements performed by the STAR Collaboration
in Au--Au collisions at $\sqrt{s_\mathrm{NN}} = 200$~GeV 
is shown in Fig.~\ref{fig:ratio} (right panel).
Since the anti-baryon-to-baryon ratio at the LHC is consistent with
unity for all transverse momenta~\cite{Aamodt:2010dx,
  PhysRevLett.109.252301}, the $\rmLambda/\Kzs$ 
 and $\bar{\rmLambda}/\Kzs$ ratios are identical and we show only the former.
The STAR $\rmLambda/\Kzs$  and $\bar{\rmLambda}/\Kzs$ ratios shown 
are constructed
by dividing the corresponding $\pT$ spectra taken 
from~\cite{Agakishiev:2011ar}. The quoted 15\% $\pT$-independent feed-down
contribution was subtracted from the $\rmLambda$ and $\bar{\rmLambda}$ spectra. 
The shape of the distributions of $\rmLambda/\Kzs$  
and $\bar{\rmLambda}/\Kzs$ are the same but they are offset by about 20\%
and have peak values around 10\% higher, and respectively lower, than the ALICE data.
This comparison between LHC and RHIC data shows that the position 
of the maximum shifts
towards higher transverse momenta as the beam energy increases.  
It is also seen that the baryon enhancement in central
nucleus--nucleus collisions at the LHC decreases less rapidly with 
$\pT$, and, at $\pT\sim 6$~GeV/$c$, it is
a factor of two higher compared with that at RHIC.

Also shown in the right panel of Fig.~\ref{fig:ratio} is a hydrodynamical model 
calculation~\cite{Song:2007fn,Song:2007ux,Song:2008si} for most central
collisions, which describes the 
$\rmLambda/\Kzs$ ratio up to $\pT$ about 2~\gmom\ rather well, 
but for higher $\pT$ progressively deviates from the data. 
Such decoupling between the calculations and measurements is already seen 
in the comparison of $\pT$-spectra~\cite{piKpPbPb:2013}. The agreement for other charged particles 
is improved when the hydrodynamical calculations are coupled to final-state 
re-scattering model~\cite{Song:2011qa}. Therefore it would be interesting to compare 
these data and their centrality evolution with such treatment. 
For higher $\pT$, a recombination model calculation~\cite{Fries:2008bh} 
is presented (Fig.~\ref{fig:ratio}, right panel). 
It approximately reproduces the shape, but overestimates the baryon 
enhancement by about 15\%. 
In the right panel of Fig.\ref{fig:ratio}, we also show a comparison of 
the EPOS model calculations~\cite{EPOS} with the current data. 
This model takes into 
account the interaction between jets and the hydrodynamically expanding
medium and arrives at a good description of the data.

In conclusion, we note that the excess of baryons at intermediate
$\pT$, exhibiting such a strong centrality dependence in \PbPb
collisions at \sqrtsNN~=~2.76~$\tev$, does not reveal
itself in pp collisions at the 
center-of-mass energy up to $\sqrt{s} = 7~\tev$.
At  \mbox{\pT$ > 7$~GeV/$c$}, the 
$\Lambda/\Kzs$ ratios measured in Pb--Pb events for different centralities all
merge together and with the dependence observed in pp collisions. This agreement between collision systems 
suggests that the relative fragmentation into 
$\rmLambda$\ and $\Kzs$\ hadrons at high $\pT$, even in central collisions, is vacuum-like and not 
modified by the medium. In future, it would be interesting to extend 
the measurements to higher transverse momenta to see whether the nuclear modification factor behaves in the same way as the one for charged particles \cite{Aamodt201130}.

As the collision energy and centrality increase,
the maximum of the $\Lambda(\bar{\rmLambda})/\Kzs$ ratio shifts towards 
higher $\pT$,
which is in qualitative agreement with the effect of increased radial flow, as predicted in~\cite{Fries:2003fr}.  
The ratio of integrated $\Lambda$ 
and \Kzs~yields does not, within uncertainties, change with centrality and is equal to that measured in pp collisions at 0.9 and 7~TeV.
This suggests that the baryon enhancement at intermediate $\pT$ is predominantly due 
to a re-distribution of baryons and mesons
over the momentum range rather than due to an additional baryon
production channel progressively opening up in more central heavy-ion collisions.
This centrality dependence may be challenging for theoretical models which try to disentangle the
quark-recombination contributions from the radial-flow 
effect and which, in addition, will need to describe other particle spectra and their $\pT$-dependent
 ratios.
 
The width of the baryon enhancement peak increases 
with the beam energy. 
However, contrary to expectations~\cite{Hwa:2008qi}, the effect 
at the LHC is still restricted to an intermediate-momentum range and is 
not observed at high $\pT$. This puts constraints on parameters of 
particle production models
involving coalescence of
quarks generated in hard parton interactions~\cite{Hwa:2012en}.   

Qualitatively, the baryon enhancement presented here as $\pT$ dependence 
of $\rmLambda/\Kzs$ ratios, is described in the low-$\pT$ region 
(below 2~\gmom) by collective hydrodynamical radial flow. In the 
high-$\pT$ region (above 7--8 GeV/c), it is very similar to pp results, 
indicating that there it is dominated by hard processes and fragmentation. 
Our data provide evidence for the need to include the effect of the 
hydrodynamical expansion of the medium formed in Pb--Pb collisions on the 
mechanisms of fragmentation and hadronization.
%

\newenvironment{acknowledgement}{\relax}{\relax}
\begin{acknowledgement}
\section*{Acknowledgements}
The ALICE collaboration would like to thank all its engineers and technicians for their invaluable contributions to the construction of the experiment and the CERN accelerator teams for the outstanding performance of the LHC complex.
\\
The ALICE collaboration acknowledges the following funding agencies for their support in building and
running the ALICE detector:
 \\
State Committee of Science,  World Federation of Scientists (WFS)
and Swiss Fonds Kidagan, Armenia,
 \\
Conselho Nacional de Desenvolvimento Cient\'{\i}fico e Tecnol\'{o}gico (CNPq), Financiadora de Estudos e Projetos (FINEP),
Funda\c{c}\~{a}o de Amparo \`{a} Pesquisa do Estado de S\~{a}o Paulo (FAPESP);
 \\
National Natural Science Foundation of China (NSFC), the Chinese Ministry of Education (CMOE)
and the Ministry of Science and Technology of China (MSTC);
 \\
Ministry of Education and Youth of the Czech Republic;
 \\
Danish Natural Science Research Council, the Carlsberg Foundation and the Danish National Research Foundation;
 \\
The European Research Council under the European Community's Seventh Framework Programme;
 \\
Helsinki Institute of Physics and the Academy of Finland;
 \\
French CNRS-IN2P3, the `Region Pays de Loire', `Region Alsace', `Region Auvergne' and CEA, France;
 \\
German BMBF and the Helmholtz Association;
\\
General Secretariat for Research and Technology, Ministry of
Development, Greece;
\\
Hungarian OTKA and National Office for Research and Technology (NKTH);
 \\
Department of Atomic Energy and Department of Science and Technology of the Government of India;
 \\
Istituto Nazionale di Fisica Nucleare (INFN) and Centro Fermi -
Museo Storico della Fisica e Centro Studi e Ricerche "Enrico
Fermi", Italy;
 \\
MEXT Grant-in-Aid for Specially Promoted Research, Ja\-pan;
 \\
Joint Institute for Nuclear Research, Dubna;
 \\
National Research Foundation of Korea (NRF);
 \\
CONACYT, DGAPA, M\'{e}xico, ALFA-EC and the EPLANET Program
(European Particle Physics Latin American Network)
 \\
Stichting voor Fundamenteel Onderzoek der Materie (FOM) and the Nederlandse Organisatie voor Wetenschappelijk Onderzoek (NWO), Netherlands;
 \\
Research Council of Norway (NFR);
 \\
Polish Ministry of Science and Higher Education;
 \\
National Authority for Scientific Research - NASR (Autoritatea Na\c{t}ional\u{a} pentru Cercetare \c{S}tiin\c{t}ific\u{a} - ANCS);
 \\
Ministry of Education and Science of Russian Federation, Russian
Academy of Sciences, Russian Federal Agency of Atomic Energy,
Russian Federal Agency for Science and Innovations and The Russian
Foundation for Basic Research;
 \\
Ministry of Education of Slovakia;
 \\
Department of Science and Technology, South Africa;
 \\
CIEMAT, EELA, Ministerio de Econom\'{i}a y Competitividad (MINECO) of Spain, Xunta de Galicia (Conseller\'{\i}a de Educaci\'{o}n),
CEA\-DEN, Cubaenerg\'{\i}a, Cuba, and IAEA (International Atomic Energy Agency);
 \\
Swedish Research Council (VR) and Knut $\&$ Alice Wallenberg
Foundation (KAW);
 \\
Ukraine Ministry of Education and Science;
 \\
United Kingdom Science and Technology Facilities Council (STFC);
 \\
The United States Department of Energy, the United States National
Science Foundation, the State of Texas, and the State of Ohio.
\end{acknowledgement}
%

\bibliographystyle{apsrev4-1} 
\bibliography{BMPR276}

\begin{thebibliography}{33}%
\makeatletter
\providecommand \@ifxundefined [1]{%
 \@ifx{#1\undefined}
}%
\providecommand \@ifnum [1]{%
 \ifnum #1\expandafter \@firstoftwo
 \else \expandafter \@secondoftwo
 \fi
}%
\providecommand \@ifx [1]{%
 \ifx #1\expandafter \@firstoftwo
 \else \expandafter \@secondoftwo
 \fi
}%
\providecommand \natexlab [1]{#1}%
\providecommand \enquote  [1]{``#1''}%
\providecommand \bibnamefont  [1]{#1}%
\providecommand \bibfnamefont [1]{#1}%
\providecommand \citenamefont [1]{#1}%
\providecommand \href@noop [0]{\@secondoftwo}%
\providecommand \href [0]{\begingroup \@sanitize@url \@href}%
\providecommand \@href[1]{\@@startlink{#1}\@@href}%
\providecommand \@@href[1]{\endgroup#1\@@endlink}%
\providecommand \@sanitize@url [0]{\catcode `\\12\catcode `\$12\catcode
  `\&12\catcode `\#12\catcode `\^12\catcode `\_12\catcode `\%12\relax}%
\providecommand \@@startlink[1]{}%
\providecommand \@@endlink[0]{}%
\providecommand \url  [0]{\begingroup\@sanitize@url \@url }%
\providecommand \@url [1]{\endgroup\@href {#1}{\urlprefix }}%
\providecommand \urlprefix  [0]{URL }%
\providecommand \Eprint [0]{\href }%
\providecommand \doibase [0]{http://dx.doi.org/}%
\providecommand \selectlanguage [0]{\@gobble}%
\providecommand \bibinfo  [0]{\@secondoftwo}%
\providecommand \bibfield  [0]{\@secondoftwo}%
\providecommand \translation [1]{[#1]}%
\providecommand \BibitemOpen [0]{}%
\providecommand \bibitemStop [0]{}%
\providecommand \bibitemNoStop [0]{.\EOS\space}%
\providecommand \EOS [0]{\spacefactor3000\relax}%
\providecommand \BibitemShut  [1]{\csname bibitem#1\endcsname}%
\let\auto@bib@innerbib\@empty
\bibitem [{\citenamefont {Adler}\ \emph {et~al.}(2003)\citenamefont {Adler}
  \emph {et~al.}}]{Adler:2003rr}%
  \BibitemOpen
  \bibfield  {author} {\bibinfo {author} {\bibfnamefont {S.~S.}\ \bibnamefont
  {Adler}} \emph {et~al.} (\bibinfo {collaboration} {PHENIX Collaboration}),\
  }\href {\doibase 10.1103/PhysRevLett.91.172301} {\bibfield  {journal}
  {\bibinfo  {journal} {Phys. Rev. Lett.}\ }\textbf {\bibinfo {volume} {91}},\
  \bibinfo {pages} {172301} (\bibinfo {year} {2003})}\BibitemShut {NoStop}%
\bibitem [{\citenamefont {Adams}\ \emph {et~al.}(2006)\citenamefont {Adams}
  \emph {et~al.}}]{Adams:2006wk}%
  \BibitemOpen
  \bibfield  {author} {\bibinfo {author} {\bibfnamefont {J.}~\bibnamefont
  {Adams}} \emph {et~al.} (\bibinfo {collaboration} {STAR Collaboration}),\
  }\href@noop {} {\  (\bibinfo {year} {2006})},\ \Eprint
  {http://arxiv.org/abs/nucl-ex/0601042} {arXiv:nucl-ex/0601042} \BibitemShut
  {NoStop}%
\bibitem [{\citenamefont {Schuster}\ \emph {et~al.}(2006)\citenamefont
  {Schuster} \emph {et~al.}}]{Schuster:2006jt}%
  \BibitemOpen
  \bibfield  {author} {\bibinfo {author} {\bibfnamefont {T.}~\bibnamefont
  {Schuster}} \emph {et~al.} (\bibinfo {collaboration} {NA49 Collaboration}),\
  }\href {\doibase 10.1088/0954-3899/32/12/S60} {\bibfield  {journal} {\bibinfo
   {journal} {J.Phys.}\ }\textbf {\bibinfo {volume} {G32}},\ \bibinfo {pages}
  {S479} (\bibinfo {year} {2006})},\ \Eprint
  {http://arxiv.org/abs/nucl-ex/0606005} {arXiv:nucl-ex/0606005} \BibitemShut
  {NoStop}%
\bibitem [{\citenamefont {Fries}\ and\ \citenamefont
  {M\"{u}ller}(2004)}]{Fries:2003fr}%
  \BibitemOpen
  \bibfield  {author} {\bibinfo {author} {\bibfnamefont {R.}~\bibnamefont
  {Fries}}\ and\ \bibinfo {author} {\bibfnamefont {B.}~\bibnamefont
  {M\"{u}ller}},\ }\href@noop {} {\bibfield  {journal} {\bibinfo  {journal}
  {Eur.\ Phys.\ J.\ C}\ }\textbf {\bibinfo {volume} {34}},\ \bibinfo {pages}
  {S279} (\bibinfo {year} {2004})}\BibitemShut {NoStop}%
\bibitem [{\citenamefont {Fries}\ \emph {et~al.}(2008)\citenamefont {Fries},
  \citenamefont {Greco},\ and\ \citenamefont {Sorensen}}]{Fries:2008bh}%
  \BibitemOpen
  \bibfield  {author} {\bibinfo {author} {\bibfnamefont {R.~J.}\ \bibnamefont
  {Fries}}, \bibinfo {author} {\bibfnamefont {V.}~\bibnamefont {Greco}}, \ and\
  \bibinfo {author} {\bibfnamefont {P.}~\bibnamefont {Sorensen}},\ }\href
  {\doibase 10.1146/annurev.nucl.58.110707.171134} {\bibfield  {journal}
  {\bibinfo  {journal} {Ann. Rev. Nucl. Part. Sci.}\ }\textbf {\bibinfo
  {volume} {58}},\ \bibinfo {pages} {177} (\bibinfo {year} {2008})},\ \Eprint
  {http://arxiv.org/abs/nucl-th/0807.4939} {arXiv:nucl-th/0807.4939}
  \BibitemShut {NoStop}%
\bibitem [{\citenamefont {Hwa}\ and\ \citenamefont {Yang}(2004)}]{Hwa:2004ng}%
  \BibitemOpen
  \bibfield  {author} {\bibinfo {author} {\bibfnamefont {R.~C.}\ \bibnamefont
  {Hwa}}\ and\ \bibinfo {author} {\bibfnamefont {C.}~\bibnamefont {Yang}},\
  }\href {\doibase 10.1103/PhysRevC.70.024905} {\bibfield  {journal} {\bibinfo
  {journal} {Phys. Rev. C}\ }\textbf {\bibinfo {volume} {70}},\ \bibinfo
  {pages} {024905} (\bibinfo {year} {2004})},\ \Eprint
  {http://arxiv.org/abs/nucl-th/0401001} {arXiv:nucl-th/0401001} \BibitemShut
  {NoStop}%
\bibitem [{\citenamefont {Hwa}(2008)}]{Hwa:2008qi}%
  \BibitemOpen
  \bibfield  {author} {\bibinfo {author} {\bibfnamefont {R.~C.}\ \bibnamefont
  {Hwa}},\ }\href@noop {} {\bibfield  {journal} {\bibinfo  {journal}
  {J.~Phys.~G.}\ }\textbf {\bibinfo {volume} {35}},\ \bibinfo {pages} {104017}
  (\bibinfo {year} {2008})},\ \Eprint {http://arxiv.org/abs/nucl-th/0804.3763}
  {arXiv:nucl-th/0804.3763} \BibitemShut {NoStop}%
\bibitem [{\citenamefont {Lamont}(2007)}]{Lamont:2007ce}%
  \BibitemOpen
  \bibfield  {author} {\bibinfo {author} {\bibfnamefont {M.}~\bibnamefont
  {Lamont}} (\bibinfo {collaboration} {STAR Collaboration}),\ }\href {\doibase
  10.1140/epjc/s10052-006-0090-9} {\bibfield  {journal} {\bibinfo  {journal}
  {Eur. Phys. J. C}\ }\textbf {\bibinfo {volume} {49}},\ \bibinfo {pages} {35}
  (\bibinfo {year} {2007})}\BibitemShut {NoStop}%
\bibitem [{\citenamefont {Aamodt}\ \emph {et~al.}(2008)\citenamefont {Aamodt}
  \emph {et~al.}}]{Aamodt:2008ul}%
  \BibitemOpen
  \bibfield  {author} {\bibinfo {author} {\bibfnamefont {K.}~\bibnamefont
  {Aamodt}} \emph {et~al.} (\bibinfo {collaboration} {ALICE Collaboration}),\
  }\href {\doibase 10.1088/1748-0221/3/08/S08002} {\bibfield  {journal}
  {\bibinfo  {journal} {JINST}\ }\textbf {\bibinfo {volume} {3}},\ \bibinfo
  {pages} {S08002} (\bibinfo {year} {2008})}\BibitemShut {NoStop}%
\bibitem [{\citenamefont {Abelev}\ \emph
  {et~al.}(2013{\natexlab{a}})\citenamefont {Abelev} \emph
  {et~al.}}]{Abelev:2013qoq}%
  \BibitemOpen
  \bibfield  {author} {\bibinfo {author} {\bibfnamefont {B.}~\bibnamefont
  {Abelev}} \emph {et~al.} (\bibinfo {collaboration} {ALICE Collaboration}),\
  }\href@noop {} {\  (\bibinfo {year} {2013}{\natexlab{a}})},\ \Eprint
  {http://arxiv.org/abs/nucl-ex/1301.4361} {arXiv:nucl-ex/1301.4361}
  \BibitemShut {NoStop}%
\bibitem [{\citenamefont {Beringer}\ \emph {et~al.}(2012)\citenamefont
  {Beringer} \emph {et~al.}}]{RPP}%
  \BibitemOpen
  \bibfield  {author} {\bibinfo {author} {\bibfnamefont {J.}~\bibnamefont
  {Beringer}} \emph {et~al.} (\bibinfo {collaboration} {Particle Data Group}),\
  }\href {\doibase 10.1103/PhysRevD.86.010001} {\bibfield  {journal} {\bibinfo
  {journal} {Phys. Rev. D}\ }\textbf {\bibinfo {volume} {86}},\ \bibinfo
  {pages} {010001} (\bibinfo {year} {2012})}\BibitemShut {NoStop}%
\bibitem [{\citenamefont {Aamodt}\ \emph
  {et~al.}(2011{\natexlab{a}})\citenamefont {Aamodt} \emph
  {et~al.}}]{Aamodt:2011zza}%
  \BibitemOpen
  \bibfield  {author} {\bibinfo {author} {\bibfnamefont {K.}~\bibnamefont
  {Aamodt}} \emph {et~al.} (\bibinfo {collaboration} {ALICE Collaboration}),\
  }\href@noop {} {\bibfield  {journal} {\bibinfo  {journal} {Eur. Phys. J. C}\
  }\textbf {\bibinfo {volume} {71}},\ \bibinfo {pages} {1594} (\bibinfo {year}
  {2011}{\natexlab{a}})}\BibitemShut {NoStop}%
\bibitem [{\citenamefont {Podolanski}\ and\ \citenamefont
  {Armenteros}(1954)}]{armenteros}%
  \BibitemOpen
  \bibfield  {author} {\bibinfo {author} {\bibfnamefont {J.}~\bibnamefont
  {Podolanski}}\ and\ \bibinfo {author} {\bibfnamefont {R.}~\bibnamefont
  {Armenteros}},\ }\href@noop {} {\bibfield  {journal} {\bibinfo  {journal}
  {Phil.~Mag.}\ }\textbf {\bibinfo {volume} {45}},\ \bibinfo {pages} {13}
  (\bibinfo {year} {1954})}\BibitemShut {NoStop}%
\bibitem [{\citenamefont {Gyulassy}\ and\ \citenamefont
  {Wang}(1994)}]{Gyulassy:1994ew}%
  \BibitemOpen
  \bibfield  {author} {\bibinfo {author} {\bibfnamefont {M.}~\bibnamefont
  {Gyulassy}}\ and\ \bibinfo {author} {\bibfnamefont {X.-N.}\ \bibnamefont
  {Wang}},\ }\href {\doibase 10.1016/0010-4655(94)90057-4} {\bibfield
  {journal} {\bibinfo  {journal} {Comp. Phys. Comm.}\ }\textbf {\bibinfo
  {volume} {83}},\ \bibinfo {pages} {307} (\bibinfo {year} {1994})},\ \Eprint
  {http://arxiv.org/abs/nucl-th/9502021} {arXiv:nucl-th/9502021} \BibitemShut
  {NoStop}%
\bibitem [{\citenamefont {Brun}\ \emph {et~al.}(1994)\citenamefont {Brun},
  \citenamefont {Carminati},\ and\ \citenamefont {Giani}}]{:1994zzo}%
  \BibitemOpen
  \bibfield  {author} {\bibinfo {author} {\bibfnamefont {R.}~\bibnamefont
  {Brun}}, \bibinfo {author} {\bibfnamefont {F.}~\bibnamefont {Carminati}}, \
  and\ \bibinfo {author} {\bibfnamefont {S.}~\bibnamefont {Giani}},\
  }\href@noop {} {\  (\bibinfo {year} {1994})},\ \bibinfo {note} {{CERN Program
  Library Long Writeup}}\BibitemShut {NoStop}%
\bibitem [{\citenamefont {ALICE}(2005)}]{aliroot-tdr}%
  \BibitemOpen
  \bibfield  {author} {\bibinfo {author} {\bibnamefont {ALICE}},\ }\href@noop
  {} {\emph {\bibinfo {title} {Technical Design Report for the ALICE Computing,
  CERN/LHCC 2005-18; AliRoot, ALICE Offline simulation, reconstruction and
  analysis framework}}},\ \bibinfo {type} {Tech. Rep.}\ (\bibinfo {year}
  {2005})\BibitemShut {NoStop}%
\bibitem [{\citenamefont {Abelev}\ \emph
  {et~al.}(2013{\natexlab{b}})\citenamefont {Abelev} \emph
  {et~al.}}]{xi_omega}%
  \BibitemOpen
  \bibfield  {author} {\bibinfo {author} {\bibfnamefont {B.}~\bibnamefont
  {Abelev}} \emph {et~al.} (\bibinfo {collaboration} {ALICE Collaboration}),\
  }\href@noop {} {\  (\bibinfo {year} {2013}{\natexlab{b}})},\ \Eprint
  {http://arxiv.org/abs/to be published} {to be published} \BibitemShut
  {NoStop}%
\bibitem [{\citenamefont {Nicassio}(2012)}]{nicassio_sqm}%
  \BibitemOpen
  \bibfield  {author} {\bibinfo {author} {\bibfnamefont {M.}~\bibnamefont
  {Nicassio}} (\bibinfo {collaboration} {for the ALICE Collaboration}),\
  }\href@noop {} {\bibfield  {journal} {\bibinfo  {journal} {Acta Phys. Polon.
  B Proc. Supp.}\ }\textbf {\bibinfo {volume} {5}},\ \bibinfo {pages} {237}
  (\bibinfo {year} {2012})}\BibitemShut {NoStop}%
\bibitem [{\citenamefont {Schnedermann}\ \emph {et~al.}(1993)\citenamefont
  {Schnedermann}, \citenamefont {Sollfrank},\ and\ \citenamefont
  {Heinz}}]{Schnedermann:1993ws}%
  \BibitemOpen
  \bibfield  {author} {\bibinfo {author} {\bibfnamefont {E.}~\bibnamefont
  {Schnedermann}}, \bibinfo {author} {\bibfnamefont {J.}~\bibnamefont
  {Sollfrank}}, \ and\ \bibinfo {author} {\bibfnamefont {U.~W.}\ \bibnamefont
  {Heinz}},\ }\href {\doibase 10.1103/PhysRevC.48.2462} {\bibfield  {journal}
  {\bibinfo  {journal} {Phys. Rev. C}\ }\textbf {\bibinfo {volume} {48}},\
  \bibinfo {pages} {2462} (\bibinfo {year} {1993})},\ \Eprint
  {http://arxiv.org/abs/nucl-th/9307020} {arXiv:nucl-th/9307020} \BibitemShut
  {NoStop}%
\bibitem [{\citenamefont {Abelev}\ \emph
  {et~al.}(2013{\natexlab{c}})\citenamefont {Abelev} \emph {et~al.}}]{pp7TeV}%
  \BibitemOpen
  \bibfield  {author} {\bibinfo {author} {\bibfnamefont {B.}~\bibnamefont
  {Abelev}} \emph {et~al.} (\bibinfo {collaboration} {ALICE Collaboration}),\
  }\href@noop {} {\  (\bibinfo {year} {2013}{\natexlab{c}})},\ \Eprint
  {http://arxiv.org/abs/to be published} {to be published} \BibitemShut
  {NoStop}%
\bibitem [{\citenamefont {Agakishiev}\ \emph {et~al.}(2012)\citenamefont
  {Agakishiev} \emph {et~al.}}]{Agakishiev:2011ar}%
  \BibitemOpen
  \bibfield  {author} {\bibinfo {author} {\bibfnamefont {G.}~\bibnamefont
  {Agakishiev}} \emph {et~al.} (\bibinfo {collaboration} {STAR
  Collaboration}),\ }\href {\doibase 10.1103/PhysRevLett.108.072301} {\bibfield
   {journal} {\bibinfo  {journal} {Phys. Rev. Lett.}\ }\textbf {\bibinfo
  {volume} {108}},\ \bibinfo {pages} {072301} (\bibinfo {year} {2012})},\
  \Eprint {http://arxiv.org/abs/nucl-ex/1107.2955} {arXiv:nucl-ex/1107.2955}
  \BibitemShut {NoStop}%
\bibitem [{\citenamefont {Song}\ and\ \citenamefont
  {Heinz}(2008{\natexlab{a}})}]{Song:2007fn}%
  \BibitemOpen
  \bibfield  {author} {\bibinfo {author} {\bibfnamefont {H.}~\bibnamefont
  {Song}}\ and\ \bibinfo {author} {\bibfnamefont {U.~W.}\ \bibnamefont
  {Heinz}},\ }\href {\doibase 10.1016/j.physletb.2007.11.019} {\bibfield
  {journal} {\bibinfo  {journal} {Phys. Lett.}\ }\textbf {\bibinfo {volume}
  {B658}},\ \bibinfo {pages} {279} (\bibinfo {year} {2008}{\natexlab{a}})},\
  \Eprint {http://arxiv.org/abs/nucl-th/0709.0742} {arXiv:nucl-th/0709.0742}
  \BibitemShut {NoStop}%
\bibitem [{\citenamefont {Song}\ and\ \citenamefont
  {Heinz}(2008{\natexlab{b}})}]{Song:2007ux}%
  \BibitemOpen
  \bibfield  {author} {\bibinfo {author} {\bibfnamefont {H.}~\bibnamefont
  {Song}}\ and\ \bibinfo {author} {\bibfnamefont {U.~W.}\ \bibnamefont
  {Heinz}},\ }\href {\doibase 10.1103/PhysRevC.77.064901} {\bibfield  {journal}
  {\bibinfo  {journal} {Phys. Rev.}\ }\textbf {\bibinfo {volume} {C77}},\
  \bibinfo {pages} {064901} (\bibinfo {year} {2008}{\natexlab{b}})},\ \Eprint
  {http://arxiv.org/abs/nucl-th/0712.3715} {arXiv:nucl-th/0712.3715}
  \BibitemShut {NoStop}%
\bibitem [{\citenamefont {Song}\ and\ \citenamefont
  {Heinz}(2008{\natexlab{c}})}]{Song:2008si}%
  \BibitemOpen
  \bibfield  {author} {\bibinfo {author} {\bibfnamefont {H.}~\bibnamefont
  {Song}}\ and\ \bibinfo {author} {\bibfnamefont {U.~W.}\ \bibnamefont
  {Heinz}},\ }\href {\doibase 10.1103/PhysRevC.78.024902} {\bibfield  {journal}
  {\bibinfo  {journal} {Phys. Rev.}\ }\textbf {\bibinfo {volume} {C78}},\
  \bibinfo {pages} {024902} (\bibinfo {year} {2008}{\natexlab{c}})},\ \Eprint
  {http://arxiv.org/abs/nucl-th/0805.1756} {arXiv:nucl-th/0805.1756}
  \BibitemShut {NoStop}%
\bibitem [{\citenamefont {Song}\ \emph {et~al.}(2011)\citenamefont {Song},
  \citenamefont {Bass},\ and\ \citenamefont {Heinz}}]{Song:2011qa}%
  \BibitemOpen
  \bibfield  {author} {\bibinfo {author} {\bibfnamefont {H.}~\bibnamefont
  {Song}}, \bibinfo {author} {\bibfnamefont {S.~A.}\ \bibnamefont {Bass}}, \
  and\ \bibinfo {author} {\bibfnamefont {U.}~\bibnamefont {Heinz}},\ }\href
  {\doibase 10.1103/PhysRevC.83.054912, 10.1103/PhysRevC.87.019902} {\bibfield
  {journal} {\bibinfo  {journal} {Phys. Rev.}\ }\textbf {\bibinfo {volume}
  {C83}},\ \bibinfo {pages} {054912} (\bibinfo {year} {2011})},\ \Eprint
  {http://arxiv.org/abs/nucl-th/1103.2380} {arXiv:nucl-th/1103.2380}
  \BibitemShut {NoStop}%
\bibitem [{\citenamefont {Werner}(2012)}]{EPOS}%
  \BibitemOpen
  \bibfield  {author} {\bibinfo {author} {\bibfnamefont {K.}~\bibnamefont
  {Werner}},\ }\href {\doibase 10.1103/PhysRevLett.109.102301} {\bibfield
  {journal} {\bibinfo  {journal} {Phys. Rev. Lett.}\ }\textbf {\bibinfo
  {volume} {109}},\ \bibinfo {pages} {102301} (\bibinfo {year}
  {2012})}\BibitemShut {NoStop}%
\bibitem [{\citenamefont {Abelev}\ \emph {et~al.}()\citenamefont {Abelev} \emph
  {et~al.}}]{piKpPbPb:2013}%
  \BibitemOpen
  \bibfield  {author} {\bibinfo {author} {\bibfnamefont {B.}~\bibnamefont
  {Abelev}} \emph {et~al.} (\bibinfo {collaboration} {ALICE Collaboration}),\
  }\href@noop {} {\ }\Eprint {http://arxiv.org/abs/hep-ex/1303.0737}
  {arXiv:hep-ex/1303.0737} \BibitemShut {NoStop}%
\bibitem [{\citenamefont {Tsallis}(1988)}]{Tsallis:1988fk}%
  \BibitemOpen
  \bibfield  {author} {\bibinfo {author} {\bibfnamefont {C.}~\bibnamefont
  {Tsallis}},\ }\href {http://dx.doi.org/10.1007/BF01016429} {\bibfield
  {journal} {\bibinfo  {journal} {J. Stat. Phys.}\ }\textbf {\bibinfo {volume}
  {52}},\ \bibinfo {pages} {479} (\bibinfo {year} {1988})}\BibitemShut
  {NoStop}%
\bibitem [{\citenamefont {Abelev}\ \emph {et~al.}(2007)\citenamefont {Abelev}
  \emph {et~al.}}]{Abelev:2006cs}%
  \BibitemOpen
  \bibfield  {author} {\bibinfo {author} {\bibfnamefont {B.}~\bibnamefont
  {Abelev}} \emph {et~al.} (\bibinfo {collaboration} {STAR Collaboration}),\
  }\href {\doibase 10.1103/PhysRevC.75.064901} {\bibfield  {journal} {\bibinfo
  {journal} {Phys.Rev.}\ }\textbf {\bibinfo {volume} {C75}},\ \bibinfo {pages}
  {064901} (\bibinfo {year} {2007})},\ \Eprint
  {http://arxiv.org/abs/nucl-ex/0607033} {arXiv:nucl-ex/0607033} \BibitemShut
  {NoStop}%
\bibitem [{\citenamefont {Aamodt}\ \emph {et~al.}(2010)\citenamefont {Aamodt}
  \emph {et~al.}}]{Aamodt:2010dx}%
  \BibitemOpen
  \bibfield  {author} {\bibinfo {author} {\bibfnamefont {K.}~\bibnamefont
  {Aamodt}} \emph {et~al.} (\bibinfo {collaboration} {ALICE Collaboration}),\
  }\href {\doibase 10.1103/PhysRevLett.105.072002} {\bibfield  {journal}
  {\bibinfo  {journal} {Phys. Rev. Lett.}\ }\textbf {\bibinfo {volume} {105}},\
  \bibinfo {pages} {072002} (\bibinfo {year} {2010})},\ \Eprint
  {http://arxiv.org/abs/nucl-ex/1006.5432} {arXiv:nucl-ex/1006.5432}
  \BibitemShut {NoStop}%
\bibitem [{\citenamefont {Abelev}\ \emph {et~al.}(2012)\citenamefont {Abelev}
  \emph {et~al.}}]{PhysRevLett.109.252301}%
  \BibitemOpen
  \bibfield  {author} {\bibinfo {author} {\bibfnamefont {B.}~\bibnamefont
  {Abelev}} \emph {et~al.} (\bibinfo {collaboration} {ALICE Collaboration}),\
  }\href {\doibase 10.1103/PhysRevLett.109.252301} {\bibfield  {journal}
  {\bibinfo  {journal} {Phys. Rev. Lett.}\ }\textbf {\bibinfo {volume} {109}},\
  \bibinfo {pages} {252301} (\bibinfo {year} {2012})}\BibitemShut {NoStop}%
\bibitem [{\citenamefont {Aamodt}\ \emph
  {et~al.}(2011{\natexlab{b}})\citenamefont {Aamodt} \emph
  {et~al.}}]{Aamodt201130}%
  \BibitemOpen
  \bibfield  {author} {\bibinfo {author} {\bibfnamefont {K.}~\bibnamefont
  {Aamodt}} \emph {et~al.} (\bibinfo {collaboration} {ALICE Collaboration}),\
  }\href {\doibase http://dx.doi.org/10.1016/j.physletb.2010.12.020} {\bibfield
   {journal} {\bibinfo  {journal} {Physics Letters B}\ }\textbf {\bibinfo
  {volume} {696}},\ \bibinfo {pages} {30 } (\bibinfo {year}
  {2011}{\natexlab{b}})}\BibitemShut {NoStop}%
\bibitem [{\citenamefont {Hwa}\ and\ \citenamefont {Zhu}()}]{Hwa:2012en}%
  \BibitemOpen
  \bibfield  {author} {\bibinfo {author} {\bibfnamefont {R.~C.}\ \bibnamefont
  {Hwa}}\ and\ \bibinfo {author} {\bibfnamefont {L.}~\bibnamefont {Zhu}},\
  }\href@noop {} {\ }\Eprint {http://arxiv.org/abs/nucl-th/1202.2091}
  {arXiv:nucl-th/1202.2091} \BibitemShut {NoStop}%
\end{thebibliography}%

\newpage
%
%
\appendix
\section{The ALICE Collaboration}
\label{app:collab}



\begingroup
\small
\begin{flushleft}
B.~Abelev\Irefn{org69}\And
J.~Adam\Irefn{org36}\And
D.~Adamov\'{a}\Irefn{org77}\And
A.M.~Adare\Irefn{org125}\And
M.M.~Aggarwal\Irefn{org81}\And
G.~Aglieri~Rinella\Irefn{org33}\And
M.~Agnello\Irefn{org87}\textsuperscript{,}\Irefn{org104}\And
A.G.~Agocs\Irefn{org124}\And
A.~Agostinelli\Irefn{org25}\And
Z.~Ahammed\Irefn{org120}
% ...
\renewcommand\labelenumi{\textsuperscript{\theenumi}~}

\section*{Affiliation notes}
\renewcommand\theenumi{\roman{enumi}}
\begin{Authlist}
\item \Adef{0}Deceased
\item \Adef{idp3706176}{Also at: M.V.Lomonosov Moscow State University, D.V.Skobeltsyn Institute of Nuclear Physics, Moscow, Russia}
\item \Adef{idp3951424}{Also at: University of Belgrade, Faculty of Physics and "Vin\v{c}a" Institute of Nuclear Sciences, Belgrade, Serbia}
\item \Adef{idp4240112}{Permanent address: Konkuk University, Seoul, Korea}
\item \Adef{idp4765344}{Also at: Institute of Theoretical Physics, University of Wroclaw, Wroclaw, Poland}
\end{Authlist}

\section*{Collaboration Institutes}
\renewcommand\theenumi{\arabic{enumi}~}
\begin{Authlist}

\item \Idef{org1}A. I. Alikhanyan National Science Laboratory (Yerevan Physics Institute) Foundation, Yerevan, Armenia
\item \Idef{org2}Benem\'{e}rita Universidad Aut\'{o}noma de Puebla, Puebla, Mexico
\item \Idef{org3}Bogolyubov Institute for Theoretical Physics, Kiev, Ukraine
\item \Idef{org4}Budker Institute for Nuclear Physics, Novosibirsk, Russia
\item \Idef{org5}California Polytechnic State University, San Luis Obispo, California, United States
\item \Idef{org6}Central China Normal University, Wuhan, China
\item \Idef{org7}Centre de Calcul de l'IN2P3, Villeurbanne, France 
\item \Idef{org8}Centro de Aplicaciones Tecnol\'{o}gicas y Desarrollo Nuclear (CEADEN), Havana, Cuba
\item \Idef{org9}Centro de Investigaciones Energ\'{e}ticas Medioambientales y Tecnol\'{o}gicas (CIEMAT), Madrid, Spain
\item \Idef{org10}Centro de Investigaci\'{o}n y de Estudios Avanzados (CINVESTAV), Mexico City and M\'{e}rida, Mexico
\item \Idef{org11}Centro Fermi - Museo Storico della Fisica e Centro Studi e Ricerche ``Enrico Fermi'', Rome, Italy
\item \Idef{org12}Chicago State University, Chicago, United States
\item \Idef{org13}Commissariat \`{a} l'Energie Atomique, IRFU, Saclay, France
\item \Idef{org14}COMSATS Institute of Information Technology (CIIT), Islamabad, Pakistan
\item \Idef{org15}Departamento de F\'{\i}sica de Part\'{\i}culas and IGFAE, Universidad de Santiago de Compostela, Santiago de Compostela, Spain
\item \Idef{org16}Department of Physics Aligarh Muslim University, Aligarh, India
\item \Idef{org17}Department of Physics and Technology, University of Bergen, Bergen, Norway
\item \Idef{org18}Department of Physics, Ohio State University, Columbus, Ohio, United States
\item \Idef{org19}Department of Physics, Sejong University, Seoul, South Korea
\item \Idef{org20}Department of Physics, University of Oslo, Oslo, Norway
\item \Idef{org21}Dipartimento di Fisica dell'Universit\`{a} and Sezione INFN, Cagliari, Italy
\item \Idef{org22}Dipartimento di Fisica dell'Universit\`{a} and Sezione INFN, Trieste, Italy
\item \Idef{org23}Dipartimento di Fisica dell'Universit\`{a} and Sezione INFN, Turin, Italy
\item \Idef{org24}Dipartimento di Fisica dell'Universit\`{a} `La Sapienza` and Sezione INFN, Rome, Italy
\item \Idef{org25}Dipartimento di Fisica e Astronomia dell'Universit\`{a} and Sezione INFN, Bologna, Italy
\item \Idef{org26}Dipartimento di Fisica e Astronomia dell'Universit\`{a} and Sezione INFN, Catania, Italy
\item \Idef{org27}Dipartimento di Fisica e Astronomia dell'Universit\`{a} and Sezione INFN, Padova, Italy
\item \Idef{org28}Dipartimento di Fisica `E.R.~Caianiello' dell'Universit\`{a} and Gruppo Collegato INFN, Salerno, Italy
\item \Idef{org29}Dipartimento di Scienze e Innovazione Tecnologica dell'Universit\`{a} del Piemonte Orientale and Gruppo Collegato INFN, Alessandria, Italy
\item \Idef{org30}Dipartimento Interateneo di Fisica `M.~Merlin' and Sezione INFN, Bari, Italy
\item \Idef{org31}Division of Experimental High Energy Physics, University of Lund, Lund, Sweden
\item \Idef{org32}Eberhard Karls Universit\"{a}t T\"{u}bingen, T\"{u}bingen, Germany
\item \Idef{org33}European Organization for Nuclear Research (CERN), Geneva, Switzerland
\item \Idef{org34}Faculty of Engineering, Bergen University College, Bergen, Norway
\item \Idef{org35}Faculty of Mathematics, Physics and Informatics, Comenius University, Bratislava, Slovakia
\item \Idef{org36}Faculty of Nuclear Sciences and Physical Engineering, Czech Technical University in Prague, Prague, Czech Republic
\item \Idef{org37}Faculty of Science, P.J.~\v{S}af\'{a}rik University, Ko\v{s}ice, Slovakia
\item \Idef{org38}Frankfurt Institute for Advanced Studies, Johann Wolfgang Goethe-Universit\"{a}t Frankfurt, Frankfurt, Germany
\item \Idef{org39}Gangneung-Wonju National University, Gangneung, South Korea
\item \Idef{org40}Helsinki Institute of Physics (HIP), Helsinki, Finland
\item \Idef{org41}Hiroshima University, Hiroshima, Japan
\item \Idef{org42}Indian Institute of Technology Bombay (IIT), Mumbai, India
\item \Idef{org43}Indian Institute of Technology Indore, India (IITI)
\item \Idef{org44}Institut de Physique Nucl\'{e}aire d'Orsay (IPNO), Universit\'{e} Paris-Sud, CNRS-IN2P3, Orsay, France
\item \Idef{org45}Institut f\"{u}r Informatik, Johann Wolfgang Goethe-Universit\"{a}t Frankfurt, Frankfurt, Germany
\item \Idef{org46}Institut f\"{u}r Kernphysik, Johann Wolfgang Goethe-Universit\"{a}t Frankfurt, Frankfurt, Germany
\item \Idef{org47}Institut f\"{u}r Kernphysik, Technische Universit\"{a}t Darmstadt, Darmstadt, Germany
\item \Idef{org48}Institut f\"{u}r Kernphysik, Westf\"{a}lische Wilhelms-Universit\"{a}t M\"{u}nster, M\"{u}nster, Germany
\item \Idef{org49}Institut Pluridisciplinaire Hubert Curien (IPHC), Universit\'{e} de Strasbourg, CNRS-IN2P3, Strasbourg, France
\item \Idef{org50}Institute for High Energy Physics, Protvino, Russia
\item \Idef{org51}Institute for Nuclear Research, Academy of Sciences, Moscow, Russia
\item \Idef{org52}Institute for Subatomic Physics of Utrecht University, Utrecht, Netherlands
\item \Idef{org53}Institute for Theoretical and Experimental Physics, Moscow, Russia
\item \Idef{org54}Institute of Experimental Physics, Slovak Academy of Sciences, Ko\v{s}ice, Slovakia
\item \Idef{org55}Institute of Physics, Academy of Sciences of the Czech Republic, Prague, Czech Republic
\item \Idef{org56}Institute of Physics, Bhubaneswar, India
\item \Idef{org57}Institute of Space Science (ISS), Bucharest, Romania
\item \Idef{org58}Instituto de Ciencias Nucleares, Universidad Nacional Aut\'{o}noma de M\'{e}xico, Mexico City, Mexico
\item \Idef{org59}Instituto de F\'{\i}sica, Universidad Nacional Aut\'{o}noma de M\'{e}xico, Mexico City, Mexico
\item \Idef{org60}iThemba LABS, National Research Foundation, Somerset West, South Africa
\item \Idef{org61}Joint Institute for Nuclear Research (JINR), Dubna, Russia
\item \Idef{org62}Korea Institute of Science and Technology Information, Daejeon, South Korea
\item \Idef{org63}KTO Karatay University, Konya, Turkey
\item \Idef{org64}Laboratoire de Physique Corpusculaire (LPC), Clermont Universit\'{e}, Universit\'{e} Blaise Pascal, CNRS--IN2P3, Clermont-Ferrand, France
\item \Idef{org65}Laboratoire de Physique Subatomique et de Cosmologie (LPSC), Universit\'{e} Joseph Fourier, CNRS-IN2P3, Institut Polytechnique de Grenoble, Grenoble, France
\item \Idef{org66}Laboratori Nazionali di Frascati, INFN, Frascati, Italy
\item \Idef{org67}Laboratori Nazionali di Legnaro, INFN, Legnaro, Italy
\item \Idef{org68}Lawrence Berkeley National Laboratory, Berkeley, California, United States
\item \Idef{org69}Lawrence Livermore National Laboratory, Livermore, California, United States
\item \Idef{org70}Moscow Engineering Physics Institute, Moscow, Russia
\item \Idef{org71}National Centre for Nuclear Studies, Warsaw, Poland
\item \Idef{org72}National Institute for Physics and Nuclear Engineering, Bucharest, Romania
\item \Idef{org73}National Institute of Science Education and Research, Bhubaneswar, India
\item \Idef{org74}Niels Bohr Institute, University of Copenhagen, Copenhagen, Denmark
\item \Idef{org75}Nikhef, National Institute for Subatomic Physics, Amsterdam, Netherlands
\item \Idef{org76}Nuclear Physics Group, STFC Daresbury Laboratory, Daresbury, United Kingdom
\item \Idef{org77}Nuclear Physics Institute, Academy of Sciences of the Czech Republic, \v{R}e\v{z} u Prahy, Czech Republic
\item \Idef{org78}Oak Ridge National Laboratory, Oak Ridge, Tennessee, United States
\item \Idef{org79}Petersburg Nuclear Physics Institute, Gatchina, Russia
\item \Idef{org80}Physics Department, Creighton University, Omaha, Nebraska, United States
\item \Idef{org81}Physics Department, Panjab University, Chandigarh, India
\item \Idef{org82}Physics Department, University of Athens, Athens, Greece
\item \Idef{org83}Physics Department, University of Cape Town, Cape Town, South Africa
\item \Idef{org84}Physics Department, University of Jammu, Jammu, India
\item \Idef{org85}Physics Department, University of Rajasthan, Jaipur, India
\item \Idef{org86}Physikalisches Institut, Ruprecht-Karls-Universit\"{a}t Heidelberg, Heidelberg, Germany
\item \Idef{org87}Politecnico di Torino, Turin, Italy
\item \Idef{org88}Purdue University, West Lafayette, Indiana, United States
\item \Idef{org89}Pusan National University, Pusan, South Korea
\item \Idef{org90}Research Division and ExtreMe Matter Institute EMMI, GSI Helmholtzzentrum f\"ur Schwerionenforschung, Darmstadt, Germany
\item \Idef{org91}Rudjer Bo\v{s}kovi\'{c} Institute, Zagreb, Croatia
\item \Idef{org92}Russian Federal Nuclear Center (VNIIEF), Sarov, Russia
\item \Idef{org93}Russian Research Centre Kurchatov Institute, Moscow, Russia
\item \Idef{org94}Saha Institute of Nuclear Physics, Kolkata, India
\item \Idef{org95}School of Physics and Astronomy, University of Birmingham, Birmingham, United Kingdom
\item \Idef{org96}Secci\'{o}n F\'{\i}sica, Departamento de Ciencias, Pontificia Universidad Cat\'{o}lica del Per\'{u}, Lima, Peru
\item \Idef{org97}Sezione INFN, Bari, Italy
\item \Idef{org98}Sezione INFN, Bologna, Italy
\item \Idef{org99}Sezione INFN, Cagliari, Italy
\item \Idef{org100}Sezione INFN, Catania, Italy
\item \Idef{org101}Sezione INFN, Padova, Italy
\item \Idef{org102}Sezione INFN, Rome, Italy
\item \Idef{org103}Sezione INFN, Trieste, Italy
\item \Idef{org104}Sezione INFN, Turin, Italy
\item \Idef{org105}SUBATECH, Ecole des Mines de Nantes, Universit\'{e} de Nantes, CNRS-IN2P3, Nantes, France
\item \Idef{org106}Suranaree University of Technology, Nakhon Ratchasima, Thailand
\item \Idef{org107}Technical University of Split FESB, Split, Croatia
\item \Idef{org108}The Henryk Niewodniczanski Institute of Nuclear Physics, Polish Academy of Sciences, Cracow, Poland
\item \Idef{org109}The University of Texas at Austin, Physics Department, Austin, TX, United States
\item \Idef{org110}Universidad Aut\'{o}noma de Sinaloa, Culiac\'{a}n, Mexico
\item \Idef{org111}Universidade de S\~{a}o Paulo (USP), S\~{a}o Paulo, Brazil
\item \Idef{org112}Universidade Estadual de Campinas (UNICAMP), Campinas, Brazil
\item \Idef{org113}University of Houston, Houston, Texas, United States
\item \Idef{org114}University of Jyv\"{a}skyl\"{a}, Jyv\"{a}skyl\"{a}, Finland
\item \Idef{org115}University of Tennessee, Knoxville, Tennessee, United States
\item \Idef{org116}University of Tokyo, Tokyo, Japan
\item \Idef{org117}University of Tsukuba, Tsukuba, Japan
\item \Idef{org118}Universit\'{e} de Lyon, Universit\'{e} Lyon 1, CNRS/IN2P3, IPN-Lyon, Villeurbanne, France
\item \Idef{org119}V.~Fock Institute for Physics, St. Petersburg State University, St. Petersburg, Russia
\item \Idef{org120}Variable Energy Cyclotron Centre, Kolkata, India
\item \Idef{org121}Vestfold University College, Tonsberg, Norway
\item \Idef{org122}Warsaw University of Technology, Warsaw, Poland
\item \Idef{org123}Wayne State University, Detroit, Michigan, United States
\item \Idef{org124}Wigner Research Centre for Physics, Hungarian Academy of Sciences, Budapest, Hungary
\item \Idef{org125}Yale University, New Haven, Connecticut, United States
\item \Idef{org126}Yonsei University, Seoul, South Korea
\item \Idef{org127}Zentrum f\"{u}r Technologietransfer und Telekommunikation (ZTT), Fachhochschule Worms, Worms, Germany
\end{Authlist}
\endgroup

%
%
\end{document}